\documentclass{elsart}
\usepackage{graphics}

\begin{document}
\begin{frontmatter}

\title{Evanescent wave diffraction of multi-level atoms }
\author{D. Gordon and C. M. Savage}
\address{Department of Physics and Theoretical Physics, \\
The Australian National University, \\
Australian Capital Territory 0200, Australia. \\
Craig.Savage@anu.edu.au}

\begin{abstract}
Diffraction of multi-level atoms by an evanescent wave reflective 
diffraction grating is modeled by numerically solving the 
time-dependent Schr\"{o}dinger equation.  We are able to explain the 
diffraction observed in experiments with metastable Neon.  This is not 
possible using a two-level atom model. The multi-level model predicts 
sensitive dependence of diffraction on the laser polarization and on 
the intensity ratio of incoming and reflected laser beams.
\end{abstract}

\end{frontmatter}

\section{Introduction}
One of the goals of atom optics is the creation of efficient analogues 
of optical elements for atoms.  Considerable progress has been made 
demonstrating mirrors, lenses, and beamsplitters \cite{AdamsAO}.  
These elements might be combined into devices such as atomic 
interferometers \cite{AdamsINT,Zeilinger,Giltner}.

Evanescent wave atomic mirrors have been demonstrated in a number of 
experiments 
\cite{Balykin,Hajnalexp,Chu,Aminoff,Seifert1,Seifert2,Feron,Stenlake}.  
A reflective diffraction grating for atoms was reported by Christ 
\emph{et al.} \cite{Christ}.  They observed diffraction of a slowed 
metastable neon beam from an evanescent optical grating formed by 
counterpropagating laser beams.  Up to 3\% of incident atoms were 
diffracted by up to 50 mrad from the main reflected beam.  This 
constitutes a large angle reflective beamsplitter.  Diffraction of 
fast metastable neon atoms has recently been observed by Brouri 
\emph{et al.} \cite{Brouri}.

Diffraction is possible because grazing incidence of the 
atomic beam produces an atomic de Broglie wavelength perpendicular to 
the grating which is comparable to the grating periodicity.  The diffraction 
angles are determined by energy and momentum conservation 
\cite{Stenlake,Hajnaltheory}.  However to determine the fraction of 
atoms which are diffracted the interaction of the atoms with the 
evanescent field must be analyzed in detail.

Two-level atom models have dominated the theoretical work on 
evanescent wave devices \cite{Hajnaltheory,Cook,Deutschmann,Murphy}.  
Although two-level models predict diffraction \cite{Deutschmann} we 
have previously reported \cite{Savage} that they cannot explain the 
\emph{particular} diffraction observed by Christ \emph{et al.} 
\cite{Christ}.  In this paper we show that multi-level atoms can 
explain the observed diffraction.  It can be understood using the 
quasipotential theory of Deutschmann \emph{et al.} \cite{Deutschmann}.  
The richer structure of the multi-level theory allows avoided 
crossings between ground state quasipotentials that are impossible in 
the two-level theory.  These produce diffraction by allowing atoms to 
exit the evanescent field in different quasipotential eigenstates than 
they entered the field \cite{Deutschmannpers}.

The multi-level model not only explains the atomic diffraction 
observed by Christ \emph{et al.} but also suggests further work.  In 
particular we find a sensitive dependence on the polarization of the 
laser beams \cite{DeutschmannMO}, and on the incoming atomic Zeeman 
level.  For pure s-polarisation,  the system can be represented as a 
series of non-interacting two-level systems and hence we do not see 
any diffraction.
\section{The model}
Our numerical model incorporates the ten magnetic sublevels of the 
3s[3/2]2 $\leftrightarrow$ 3p$'$[3/2]2 transition of metastable neon.  
By numerically solving the time dependent Schr\"{o}dinger equation we 
find diffracted beam fractions consistent with the observations of 
Christ \emph{et al.} \cite{Christ}. 

\begin{figure}[ht]
\includegraphics{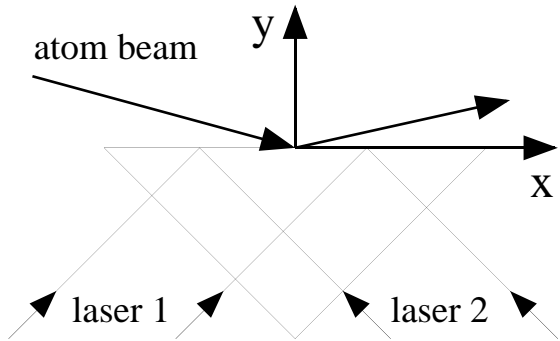}
\caption{The geometry of the lasers and atomic beam in the evanescent field 
diffraction experiment. The atomic beam angles are exaggerated for clarity.}
\protect\label{schematic}
\end{figure}

The geometry of the diffraction experiment is shown in Fig.\ 
\ref{schematic}.  The direction perpendicular to the quartz surface is 
$y$ and the direction parallel to the surface $x$.  The origin of the 
$y$ coordinate is at the interface and the positive direction is into 
the vacuum.  We assume the atom is initially propagating towards the 
interface in the $xy$ plane with a positive $x$-component of velocity.

The modulus of the evanescent field wavevector $Q$ and its inverse 
decay length $q$ are \cite{Born}
\begin{equation}
	Q = k n_{q} \sin \theta , \quad
	q = k \sqrt{(n _{q} \sin \theta)^{2}-1} ,
	\label{wavevec}
\end{equation}
where $k$ is the free field wavevector modulus, assumed to be the 
same for each laser, $n_{q}$ is the refractive index of the quartz, 
and $\theta$ is the angle of incidence of the lasers with 
the quartz surface. The values of these parameters for the experiment 
we shall model are given in Table 1.

\begin{table}
\caption{Parameters used in experiment of Christ \textit{et al}.\ 
\protect\cite{Christ}. }
\begin{tabular}{cc} 
\hline
Parameter & Value  \\ 
\hline
Atom                 & Ne* \\
Atom mass            & $3.3 \times 10^{-26}$ kg \\
Atomic velocity      & 25 ms$^{-1}$ \\ 
Angle of incidence   & 36 mrad \\
Laser detunings      & 900 MHz \\
Transition wavelength& 594.5 nm \\
$k$                  & $1.058 \times 10^7$ m$^{-1}$ \\
$Q$                  & $1.10 \times 10^7$ m$^{-1}$ \\
$q$                  & $2.72 \times 10^6$ m$^{-1}$ \\
\hline
\end{tabular}
\end{table}

We assume that the atoms are initially in an eigenstate $| k_{0x} 
\rangle$ of the $x$-component of their momentum with eigenvalue $\hbar 
k_{0x}$.  Then because the photon momentum of the copropagating 
(counterpropagating) field is $\hbar Q$ ($-\hbar Q$) the $x$-component 
of the atomic momentum is restricted to the eigenvalues $\hbar (k_{0x} 
+ nQ)$, with $n$ any integer.  We denote the corresponding set of 
allowed centre-of-mass $x$-momentum eigenstates by $ \{|n \rangle, n 
\in \mathrm{integers} \} $.  A ground (excited) state atom has $n$ 
even (odd).

\begin{figure}[ht]
\includegraphics{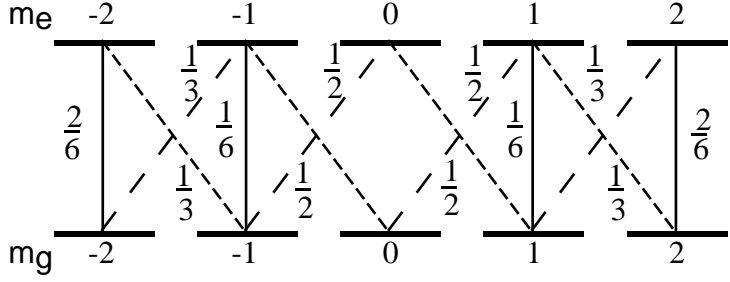}
\caption{Schematic diagram of the atomic 
level structure used in the model.  The squares of the Clebsch-Gordon 
coefficients $C^{p}_{m_g,m_e}$ for the transitions are shown.}
\protect\label{levels}
\end{figure}

The 3s[3/2]2 $\leftrightarrow$ 3p$'$[3/2]2 transition of metastable 
neon is between states with total angular momentum J=2.  Hence there 
are five ground $| m_{g} \rangle$, and five excited $| m_{e} \rangle$, 
sublevels labeled by the magnetic quantum numbers $m_{g/e} \in 
\{-2,-1,0,1,2\}$, Fig.\ \ref{levels}.
The \emph{discrete} basis states for the $x$-component of momentum and 
the internal state of the atom are therefore
\begin{equation}
	|n,m \rangle \equiv |n \rangle \otimes |m \rangle .
	\label{basis}
\end{equation}
Given our initial condition of ground state atoms the magnetic quantum 
number $m$ is $m_{g}$ ($m_{e}$) if $n$ is even (odd). The atomic motion in 
the $y$-direction, perpendicular to the quartz surface, is represented 
by one coordinate basis wavefunction $\Psi_{n,m}(y,t)$ for each 
discrete basis state $|n,m \rangle$.  We use the following ansatz for 
the complete state of the atom
\begin{equation}
	| \Psi(t) \rangle = 
	\exp(-ik_{0x}^{2}t/2M\hbar)  
	\sum_{m,n} \Psi_{n,m}(y,t) |n,m \rangle ,
	\label{ansatz}
\end{equation}
where $M$ is the atomic mass. The exponential prefactor 
accounts explicitly for the initial atomic kinetic energy in the 
$x$-direction.

The Hamiltonian $H$ for the system can be divided into two parts: the 
$y$-component of the kinetic energy $p_{y}^{2} / 2M$, with continuous 
eigenvalues, and the rest $\{ V + p_{x}^{2} / 2M \}$, which has discrete 
eigenvalues,
\begin{equation}
	H = \frac{p_{y}^{2}}{2M} + 
	\left\{ \frac{p_{x}^{2}}{2M} + V \right\} .
	\label{hamiltonian}
\end{equation}
$p_{y}$ ($p_{x}$) is the $y$-component ($x$-component) of the atomic 
momentum perpendicular (parallel) to the quartz surface.  $V$ is the 
sum of the atom's internal energy and the electric dipole interaction 
energy between the evanescent field and the atomic transition 
$H_{\mathrm{ED}}$,
\begin{equation}
		V = 
	\sum_{m,n(\mathrm{odd})} \hbar \Delta_a
	|m,n\rangle  \langle m,n| + H_{\mathrm{ED}} ,
	\label{discreteham}
\end{equation}
This Hamiltonian is in an interaction picture 
with the atomic dipoles rotating at the average of the laser 
frequencies $\bar{\omega} = (\omega_{1}+\omega_{2})/2$, where 
$\omega_{1} (\omega_{2})$ is the frequency of the laser copropagating 
(counterpropagating) with the atoms.  These frequencies can be 
different, as in the experiment of Stenlake \emph{et al.} 
\cite{Stenlake}.  In that case they must be sufficiently similar that 
the approximation of equal photon momentum magnitudes holds.  The 
atomic detuning $\Delta_a$ is the difference between the (degenerate) 
atomic transition frequencies and $\bar{\omega}$.  The dipole 
interaction energy is
\begin{equation}
		H_{\mathrm{ED}} = \mathbf{d}^{+} \cdot 
		\mathbf{E}^{+} + \mathrm{H.c.}
	\label{hamed}
\end{equation}
where H.c.\ means Hermitean conjugate and $\mathbf{d}^{+}$ is the 
positive frequency part of the transition electric dipole moment 
operator $\mathbf{d}$
\begin{equation}
	\mathbf{d}^{+} = 
    \sum_{m_{g},m_{e} = -J}^{J} |m_e\rangle 
     \langle m_e| {\bf d} |m_g\rangle  \langle m_g| .
	\label{dipole+}
\end{equation}
$\mathbf{E}^{+}$ is the positive frequency part of the total electric 
field
\begin{equation}
	\mathbf{E}^{+} = 
	\mbox{\boldmath $\varepsilon$}_1
	E_1 (y,t)  \exp(iQ x) + 
	\mbox{\boldmath $\varepsilon$}_2 
	E_2 (y,t)  \exp(-i Q x) ,
	\label{field+}
\end{equation}
\begin{equation}
	E_{i}(y,t) = \exp(-q y) \exp( -i \Delta_{d} t) E_{i0}.
	\label{edefn}
\end{equation}
$E_{10}$ ($E_{20}$) is the amplitude of the copropagating 
(counterpropagating) field at the quartz surface, and is assumed real 
for convenience.  The $\mbox{\boldmath $\varepsilon$}_{i}$ are the 
corresponding field polarization vectors.  $\Delta_{d} = 
(\omega_{1}-\omega_{2})/2$ is half the frequency difference between 
the two lasers.  In the examples we consider in the next section 
$\Delta_{d} = 0$.

We take the $z$ direction as the quantization axis and work with a 
spherical basis of polarization vectors $\{ \mathbf{u}_{\pm}, 
\mathbf{u}_{0} \}$ defined in terms of the Cartesian unit vectors by
\begin{equation}
	\mathbf{u}_{\pm} = \frac{1}{\sqrt{2}} 
	( \mathbf{u}_{x} \pm i\mathbf{u}_{y} ) , \quad
	\mathbf{u}_{0} =\mathbf{u}_{z} .
	\label{polbasis}
\end{equation}
The Wigner-Eckart theorem \cite{Merzbacher} allows us to express the 
dipole matrix elements in terms of Clebsch-Gordon coefficients 
$C^{p}_{m_g,m_e}$ and a reduced matrix element $\mathcal{D}$
\begin{equation}
	\langle m_e| {\bf d} \cdot \mathbf{u}_{p} |m_g\rangle =
	C^{p}_{m_g,m_e} \mathcal{D} ,
	\label{matel}
\end{equation}
where $p$ is $0, \pm1$ for $\pi, \sigma_{\pm}$ transitions.
The values of the non-zero Clebsch-Gordon coefficients are indicated 
in Fig.\ \ref{levels}.  The dipole interaction Hamiltonian 
Eq.\ (\ref{hamed}) then becomes
\begin{eqnarray}
	H_{\mathrm{ED}} &=& \mathcal{D} \sum_{p,m_g,m_e} \left\{
	E_1( y,t) \sum_{n(even)} 
	C^{p}_{m_g,m_e} 
	\varepsilon_{1,p} |m_e,n+1\rangle \langle m_g,n |
	\right.
	\nonumber \\
	& & \left. + E_2( y,t) \sum_{n(odd)} 
	C^{p}_{m_g,m_e} 
	\varepsilon_{2,p} |m_g,n+1\rangle \langle m_e,n |
	\right\} +\mathrm{H.c.}
	\label{reduced}
\end{eqnarray}

The $\varepsilon_{1/2,p} = \mbox{\boldmath $\varepsilon$}_{1/2} \cdot 
\mathbf{u}_{p}$ are the spherical polarization components of 
the two lasers.

Substituting the ansatz Eq.\ (\ref{ansatz}) into the Schr\"{o}dinger 
equation corresponding to the Hamiltonian Eq.\ (\ref{hamiltonian}) 
gives the set of coupled Schr\"{o}dinger equations
\begin{eqnarray}
i\partial_t\Psi_{n,m_{e/g}} &=& \left[ \frac{p_y^2}{2M\hbar} +
S_n\right]\Psi_{n,m_{e/g}} \nonumber \\
&& + \sum_{n',m_{e/g}'} 
\langle n,m _{e/g} | V |n',m_{e/g}'\rangle \Psi_{n',m_{e/g}'} \quad,
	\label{coupledse}
\end{eqnarray}
\begin{equation}
S_n \equiv \frac{\hbar}{2M}(2k_{0,x}nQ + n^2Q^2).
\end{equation}
We have solved this set of partial differential equations numerically 
by the well known split operator method \cite{Savage}.  Our numerical 
solutions were checked for accuracy by demanding that changing either 
the time-step or the spatial grid size did not significantly alter 
the final result.

\section{Diffraction}
Before presenting numerical solutions of the Schr\"{o}dinger equations 
we consider the quasipotentials of the Hamiltonian Eq.\ 
(\ref{hamiltonian}) \cite{Deutschmann}.  These are the eigenvalues of 
that part of the Hamiltonian having discrete eigenvalues, namely $\{ V 
+ p_{x}^{2} / 2M \}$ Eq.\ (\ref{discreteham}).  Since $V$ includes the 
interaction with the evanescent field the quasipotentials are a 
function of $y$, the distance from the quartz surface.  The 
eigenstates corresponding to the quasipotentials may be thought of as 
atomic states doubly dressed by the two evanescent fields.

The eigenstates of $V$ corresponding to the quasipotentials are 
adiabatically followed by atoms moving sufficiently slowly towards the 
surface, except near avoided crossings of the quasipotentials.  
Hence they behave like actual potentials, for example slowing atoms 
down as they climb them, which is the origin of atomic reflection 
\cite{SavageAJP}.  At the avoided crossings non-adiabatic transitions 
between the quasipotentials may occur.  These make diffraction 
possible, since the atoms may leave the evanescent field in a 
different superposition of quasipotential eigenstates than that in 
which they entered, and different eigenstates may have different momenta.

A two-level atom model allows diffraction for certain ranges of 
parameters.  In particular the diffraction considered by Deutschmann 
\textit{et al}.\ \cite{Deutschmann} occurred for much lower atomic 
detunings than were used in the experiment of Christ \textit{et al}.\ 
\cite{Christ}.  The lower detunings would have produced unacceptable 
levels of atomic excitation and hence of spontaneous emission.

\begin{figure}[ht]
\includegraphics{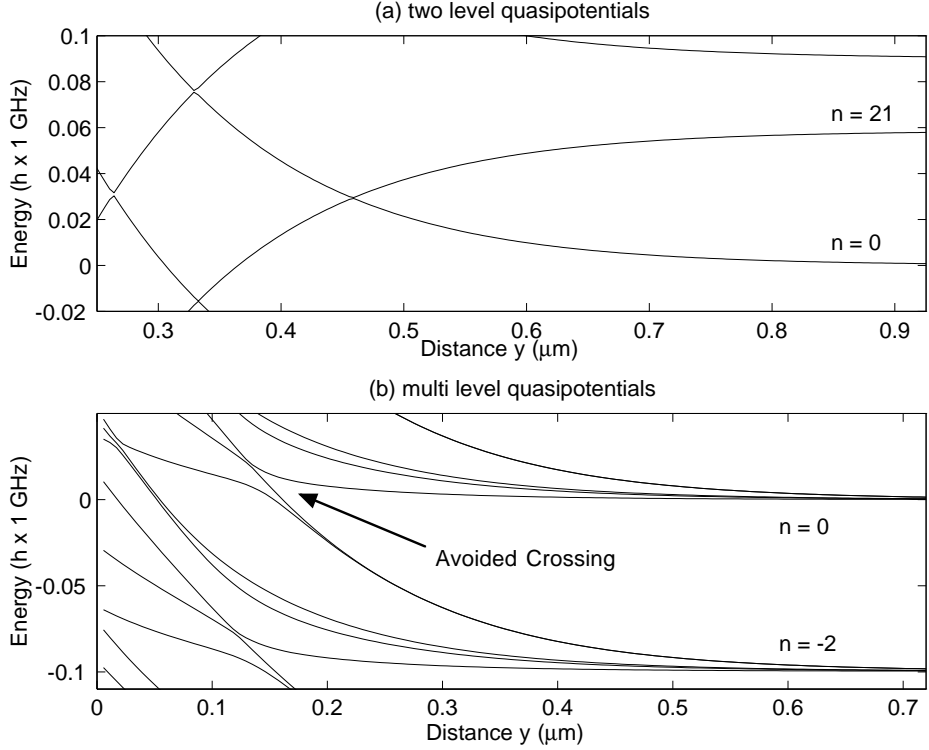} \caption{Quasipotentials versus perpendicular distance 
from the quartz surface, for the parameters of Table 1.  
(a) Two-level quasipotentials.  (b) Multi-level 
quasipotentials for a laser polarization angle of $5^{\circ}$ away 
from p-polarization. }
\protect\label{QP}
\end{figure}

Since excited states spontaneously emit, diffraction 
into ground states, with even $n$, is of most interest.  The optimum 
probability for the $n=\pm 2$ diffraction orders was estimated to be 
about 6\% by Deutschmann \textit{et al}.\ \cite{Deutschmann}.  This is 
because at least four avoided crossings are required, with an optimal 
transfer probability of 50\% at each, and $0.5^{4} \approx 0.06$.  The 
quasipotentials for a two-level model of the experiment are shown in 
Fig.\ \ref{QP}(a).  Note that the incoming $n=0$ quasipotential only 
has avoided crossings with high order quasipotentials, the first being 
with the $n=21$ quasipotential \cite{Deutschmannpers}.  The transition 
between the corresponding eigenstates would involve a 21 photon 
process and hence be very weak, giving negligible $n=-2$ order 
diffraction.  This is inconsistent with the observations of Christ 
\textit{et al}.\ \cite{Christ}.

The quasipotentials for the multi-level atom model of the experiment 
are shown in Fig.\ \ref{QP}(b).  The multi-level model makes 
possible Raman transitions between the ground Zeeman levels.  Hence 
there is now an avoided crossing between $n=0$ and $n=-2$ 
quasipotentials,  so diffraction into an $n=-2$ eigenstate is 
expected. 

\begin{figure}
\includegraphics{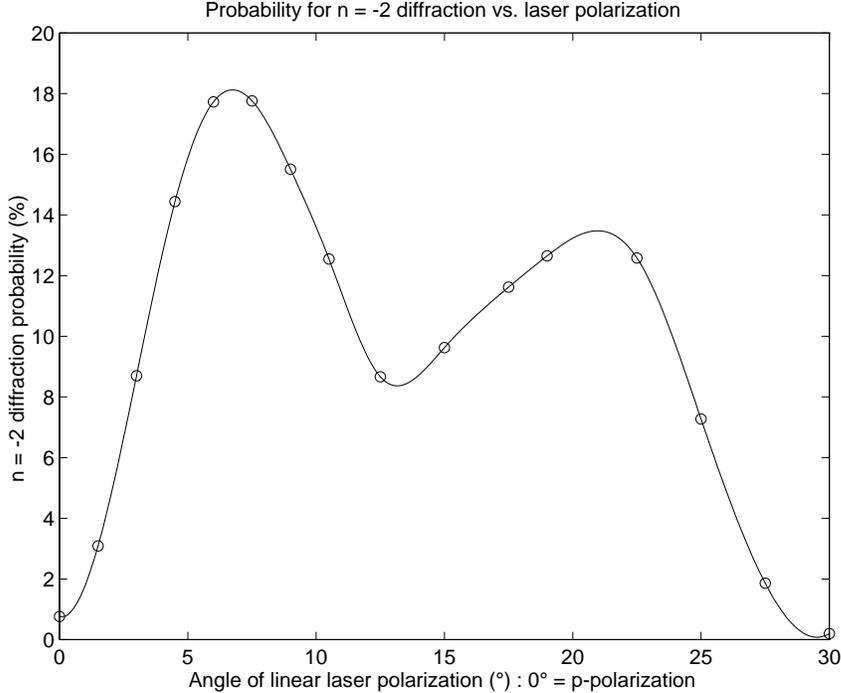} \caption{Percentage $n = -2$ order diffraction versus polarization 
angle for an incoming $m_{g} = 2$ atom.  The percentage of diffraction is plotted versus the linear 
polarization angle, in degrees, of the two laser beams from pure 
p-polarization, $0^{\circ}$.  The solid curve is a fit to the 
calculated points.  Other parameters are as in Table 1.}
\protect\label{diffvpol}
\end{figure}

The experiment of Christ \emph{et al.} \cite{Christ} that we have
modelled reported up to 3\% diffraction.  The ratio of the intensities of 
the copropagating to counterpropagating waves was fixed at 1.64.  
To calculate the diffracted fraction we numerically solved the 
Schr\"{o}dinger equations (\ref{coupledse}), as described in the 
previous section.  Our results are quite sensitive to the 
polarizations of the two laser beams \cite{DeutschmannMO}. Fig. 
\ref{diffvpol} shows the percentage of diffraction into the $n=-2$ 
order as a function of the polarization angle of the lasers. 
The results in this figure were calculated for an equal mixture of 
magnetic sublevels in the incoming atomic beam.

We did not calculate the diffraction from all initial $m_{g}$ states 
for all polarization angles.  This was because each point on Fig.  4 
took approximately 30 minutes to compute on a VPP300 supercomputer.  
Typically the spatial grid in the $y$ direction had 2048 elements and 
13 $n$ states, $n=-6$ to $n=6$ were used.  The integration over  
1.6 $\mu$s was performed with 0.2 ns timesteps.  These numerical 
parameters were varied to ensure the insensitivity of the solutions.

However we did calculate for all $m_{g}$ states for several 
representative polarisation angles,  and in all cases it was found 
that they produced amounts of diffraction bearing a fixed ratio to 
each other.  This is because only one of the $n = 0$ quasipotentials 
is involved in the diffraction process,  and the various $m_{g}$ 
states enter this quasipotential in fixed ratios.  This observation 
allowed us to deduce the overall diffraction for a mixture of 
magnetic sublevels.

We have also modelled the conditions which gave maximum diffraction 
according to unpublished experimental data \cite{Deutschmannpers}: The 
polarization of the copropagating beam rotated by 5 
degrees from perfect p-polarization (electric field in 
the plane of incidence) and the counterpropagating beam 
rotated 15 degrees from perfect p-polarization 
\cite{Deutschmannpers}.  Despite uncertainty as to the accuracy of 
these figures,  we found that these parameters gave a large degree of 
diffraction (about 14\%).  Furthermore,  a ratio of 1.64 between the 
co-propagating and counterpropagating beams was cited as an important 
condition for diffraction to occur \cite{Christ}.  Computationally,  we found 
that this condition produced a local maximum in the amount of 
diffraction.

These results demonstrate that a multi-level model is able to account 
for the diffraction observed by Christ \emph{et al.} \cite{Christ}. 
The difference between the diffraction we find (14\% for 
experimental parameters) and that observed, 3\%, is reasonable 
given the non-ideal aspects of the 
experiment and the lack of delicate control over the parameters.   
 
\begin{figure}
\includegraphics{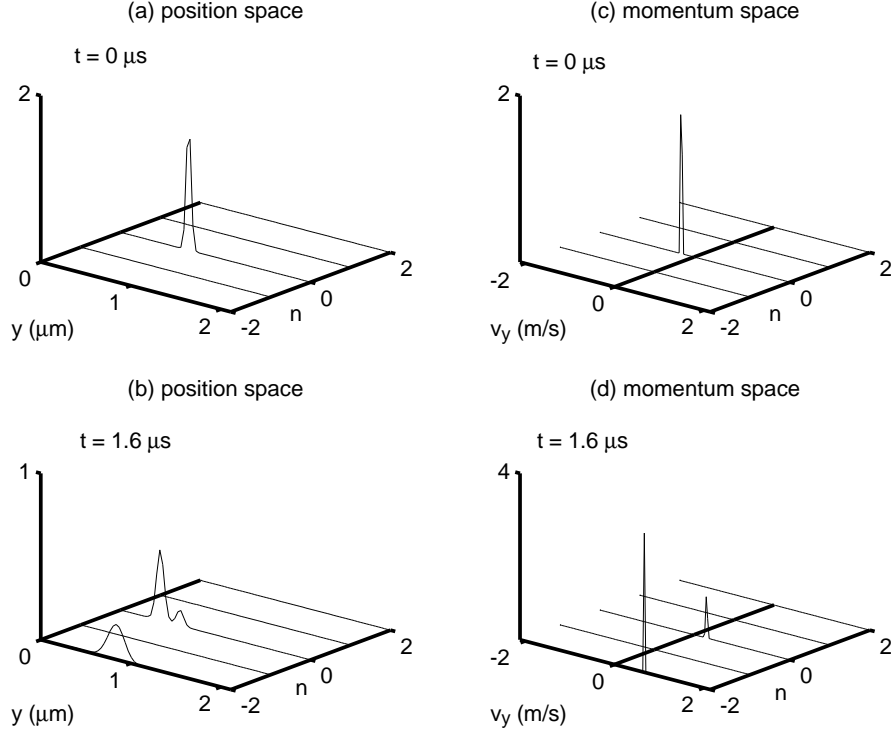}
\caption{Probability densities in position and momentum spaces for an 
initial $m_{g}= 2$ state atom.  The probability density $P_{n} = \sum_{m} 
|\Psi_{n,m}|^{2}$ is plotted against the diffraction order $n$ and 
either the distance from the glass in microns (a) and (b), or the 
velocity perpendicular to the glass (c) and (d).  In each case the top 
figure is the initial condition and the bottom figure is 1.6 $\mu$s 
later, after reflection.  There is no probability in the orders not 
shown.  Parameters are as in Table 1 with laser polarizations of 
$5^{\circ}$ away from p-polarization.}
\protect\label{wf}
\end{figure}

One of the advantages of numerically solving the time-dependent 
Schr\"{o}dinger equation is that ``movies'' of the wavefunction evolution 
are available.  The wavefunction can be visualized in either 
coordinate or momentum space.  Fig.\ \ref{wf} shows an initial 
$m_{g}=2, n=0$ probability density evolving to produce 
$n= -2$ order diffraction.

\section{Conclusion}
We have quantitatively modeled, from first principles, the reflection 
grating atomic diffraction experiment of Christ \emph{et al.} 
\cite{Christ}.  Our results are consistent with the levels of 
diffraction observed in that experiment.  The model predicts strong 
dependence on the polarization of the copropagating and 
counterpropagating laser beams.  This is physically reasonable since 
the atomic Zeeman structure makes the shifts of the various atomic 
transitions polarization dependent.  There is some evidence for this 
effect in the experiment \cite{Deutschmannpers}.
 
Computational modelling of experiments, such as we have reported, is 
particularly useful if it can suggest new experiments.  Our results 
show that control of the polarization of both laser beams is crucial 
in reflection grating atomic diffraction experiments.  Furthermore we 
found that under the conditions of the experiment of Christ \emph{et 
al.} \cite{Christ} the Zeeman $m_{g}= 2$ state produced most of the 
$n=-2$ order diffraction.  Hence optical pumping into this state could 
potentially increase the diffraction.

Our model is easily adapted to other atoms and we plan to use it to 
model a Caesium evanescent wave diffraction experiment currently 
underway in our group.

\section*{Acknowledgments} 

We are particularly indebted to R. Deutschmann and M. Schiffer for 
correspondence and unpublished data concerning their experiment. 
We also acknowledge discussions with the ANU atom optics 
group, especially, I.\ Littler and J.\ Eschner.  The computations were 
performed at the Australian National University Supercomputer 
Facility.


\end{document}